\documentstyle[12pt,a4,psfig]{article}
\begin{document}
\baselineskip 6.5mm

\begin{flushright}
{\tt hep-th/0105044}
\end{flushright}
\vspace{5mm}

\begin{center}
\large \bf Dielectric-branes in Non-supersymmetric 
SO(3)-invariant Perturbation of Three-dimensional ${\cal N}=8$ Yang-Mills 
Theory
\end{center}

\vspace{5mm}

\begin{center}
Changhyun Ahn \footnote{Email address: ahn@knu.ac.kr}~~~and~~~
Taichi Itoh \footnote{Email address: taichi@knu.ac.kr}
\end{center}

\begin{center} 
\it Department of Physics, Kyungpook National University, 
Taegu 702-701, Korea\\
\end{center}

\vspace{5mm}

\begin{center}
Abstract
\end{center}

We study non-supersymmetric $SO(3)$-invariant deformations of $d=3$, 
${\cal N}=8$ super Yang-Mills theory and their type IIA string theory dual.  
By adding both gaugino mass and scalar mass, dielectric D4-brane potential
coincides with D5-brane potential in type IIB theory.
We find the region of parameter space where the non-supersymmetric vacuum
is described by stable dielectric NS5-branes. By considering the generalized
action for NS5-branes in the presence of D4-flux, we also analyze the 
properties of dielectric NS5-branes.

\vspace{10mm}
\noindent

\newpage


\noindent
{\bf 1. Introduction}\\

The Anti de Sitter(AdS)/Conformal Field Theory(CFT) 
correspondence (for a review, see \cite{agmoo})
enables us to study not confining theories but conformal ${\cal N}=4$
gauge theories that are dual to type IIB string theory on $AdS_5 \times
{\bf S}^5$. In order to understand the former, one can perturb
by adding mass terms preserving some or none of supersymmetry and gets 
a confining gauge theory. It is known that in \cite{polchinski}, they made
a proposal for the dual supergravity description of a 
four-dimensional confining gauge theory by adding finite mass terms
to ${\cal N}=4$ Yang-Mills theory and computed the linearized perturbed
background by the presence of non-vanishing boundary conditions on
the magnetic three-form. In each of the many vacua, D3-branes were replaced by
several five-branes through Myers' dielectric effect \cite{myers}.    
It turned out that as long as the ratio of five- and three-brane charge
densities is very small, the solutions are good near five-brane action minima. 

Motivated by the work of \cite{polchinski}, the dual M-theory 
description of a three-dimensional theory living on M2-branes by adding 
fermion mass terms was found \cite{bena1}. 
Similarly, the nonsingular string theory duals corresponding to a perturbed 
three-dimensional gauge theory on D2-branes was obtained by the polarization 
of D2-branes into D4-branes and NS5-branes \cite{bena2}. 
Moreover, the dual string theory of oblique vacua in the perturbed 
three-dimensional gauge theory corresponded to the polarization of D2-branes
into NS5-branes with D4-brane charge \cite{bena3}. 
In \cite{zamora}, $SO(3)$-invariant deformations of four-dimensional
${\cal N}=4$ gauge theory within the context of \cite{polchinski} 
was studied and the non-supersymmetric vacuum is described by stable 
dielectric five-branes. 

In this paper, we consider perturbed three-dimensional gauge theory
living on D2-branes by adding both gaugino mass term and the scalar terms and
construct dual string theory corresponding to this $SO(3)$-invariant 
non-supersymmetric deformation of $d=3$, ${\cal N}=8$ theory.
In section 2, we review three-dimensional Yang-Mills theory and how
its $SO(3)$-invariant perturbations appear. In section 3, for given 
${\cal N}=2$ supersymmetric gauge theory perturbed by
three fermion masses, we go one step further by inserting 
gaugino mass and scalar mass which are $\bf 35$ and $\bf 27$ of 
$SO(7)_R$ symmetry, respectively. It turns out it is exactly same 
form of the one of type IIB theory described by D3/D5 potential.
Similarly, in section 4, we consider the scalar mass $\bf 27$ of
$SO(7)_R$ symmetry, modify ${\cal N}=2$ dielectric NS5-brane action
and study its phase diagram. In section 5, we do the same analysis for
generalized NS5-brane action. In section 6, we make our conclusions and
future directions. \\

\noindent
{\bf 2. The SO(3)-invariant perturbations in type IIA theory}\\

We start from preliminaries about the type IIA D2-brane solution and 
its $SO(3)$-invariant perturbation both in the bulk and in the boundary. 
The unperturbed space-time generated by $N$ coincident D2-branes 
is obtained as \cite{horowitz,itzhaki}
\begin{equation}
ds^2 = Z^{-1/2}\eta_{\mu\nu}dx^\mu dx^\nu +Z^{1/2}\delta_{mn}dx^m dx^n,
\end{equation}
with the R-R three-form potential
$$
C^0_3 = -\frac{1}{g_s Z}\,dx^0 \wedge dx^1 \wedge dx^2,
$$
where $\mu,\nu=0,1,2$, $m,n=3,\dots,9$, and $g_s$ is the string coupling 
which is related to the dilaton field $\Phi$ through $e^\Phi =g_s Z^{1/4}$. 
The warp factor $Z$ is given by a harmonic function
$$
Z=\frac{R^5}{r^5},\quad R^5 =6\pi^2 Ng_s \alpha^{\prime\,5/2}.
$$
The dielectric D4- and NS5-brane configurations are obtained by perturbing 
the type IIA D2-brane solution with linearized $H_3 =dB_2$, $F^1_4 =dC^1_3$ 
field strengths which transform as $\bf 35$ of $SO(7)_R$ rotation group in 
transverse seven-dimensions. 

The gauge theory living on $N$ coincident D2-branes is an ${\cal N}=8$ super 
Yang-Mills \cite{seiberg}. An ${\cal N}=8$ gauge multiplet consists of 
a gauge field $A_\mu$, eight real fermions $\{\lambda_1,\dots,\lambda_8\}$ 
which are in $\bf 8$ spinor representation of $SO(7)_R$ $R$-symmetry, 
and seven real scalars $\{\chi_1,\dots,\chi_7\}$ which are in $\bf 7$ vector 
representation of the $SO(7)_R$. 
The eight fermions are cast into four complex fermions 
$$
\Lambda_1 =\lambda_1 +i\lambda_2,\quad 
\Lambda_2 =\lambda_3 +i\lambda_4,\quad
\Lambda_3 =\lambda_5 +i\lambda_6,\quad
\Lambda_4 =\lambda_7 +i\lambda_8,
$$
which transform as $\bf 4$ of $SU(4)\subset SO(7)_R$, whereas the seven 
scalars are divided into six scalars of $\bf 6$ of $SU(4)$ and an 
$SU(4)$-singlet real scalar. The six scalars are cast into three complex 
scalars
$$
\phi_1 =\chi_1 +i\chi_2,\quad 
\phi_2 =\chi_3 +i\chi_4,\quad
\phi_3 =\chi_5 +i\chi_6,
$$
which are combined together with $\Lambda_1$, $\Lambda_2$, $\Lambda_3$ into 
three ${\cal N}=2$ hypermultiplets $\{\phi_i,\Lambda_i\}$, $i=1,2,3$, 
transforming as $\bf 3$ of $SU(3)\subset SU(4)$. The $SO(3)$ group considered 
in this paper is a real subgroup of the $SU(3)$. 
The gauge boson $A_\mu$ and the singlet scalar $\chi_7$ are made up into 
an ${\cal N}=2$ gauge multiplet $\{A_\mu,\chi_7,\Lambda_4\}$ together with 
$\Lambda_4$ including a gaugino field $\lambda_8$. 

Both $d=3$, ${\cal N}=8$ and $d=4$, ${\cal N}=4$ super Yang-Mills theories 
share sixteen supercharges \cite{seiberg}. The ${\cal N}=2$ gauge (hyper) 
multiplet in three-dimensions correspond to the ${\cal N}=1$ gauge (chiral) 
multiplet through the dimensional reduction. The $SU(4)$ subgroup considered 
above is nothing but the $R$-symmetry of $d=4$, ${\cal N}=4$ super Yang-Mills. 
This implies that the $d=3$, ${\cal N}=2$ gauge theory obtained by giving 
masses $m_1$, $m_2$, $m_3$ to the three hypermultiplets corresponds to 
the $d=4$, ${\cal N}=1$ gauge theory \cite{vafa,donagi} considered in 
\cite{polchinski}. The ${\cal N}=2$ fermion mass terms appear in the 
Lagrangian
\begin{equation}
\Delta L ={\rm Re}\left( m_1 \Lambda_1^2+m_2 \Lambda_2^2+m_3 \Lambda_3^2
+m_4 \Lambda_4^2 \right),\label{fermion}
\end{equation}
where the gaugino mass $m_4$ must be zero to obtain the ${\cal N}=2$ gauge 
theory otherwise the gauge theory becomes non-supersymmetric. 
If we set $m_1 =m_2 =m_3 \equiv m$, the mass perturbation becomes 
$SO(3)$-invariant. Both real and imaginary parts of 
$\{\Lambda_1,\Lambda_2,\Lambda_3\}$ are $\bf 3$ of $SO(3)$, while 
$\Lambda_4$ is an $SO(3)$-singlet. 

In order to make contrast of the type IIA $SO(3)$-invariant perturbation 
with the one in type IIB case, it may be useful to show the branching 
rules of $SO(7)_R \to SU(4)_R$:
\begin{eqnarray}
{\bf 8} &\to& {\bf 4}+\overline{\bf 4},\nonumber\\
{\bf 7} &\to& {\bf 6}+{\bf 1},\nonumber\\
{\bf 35} &\to& {\bf 10}+\overline{\bf 10} +{\bf 15},\nonumber\\
{\bf 27} &\to& {\bf 20}^\prime +{\bf 6}+{\bf 1}.
\end{eqnarray}
Each of the $H_3$ and $F^1_4$ perturbations corresponds to $\bf 35$ of 
$SO(7)_R$, that is, a traceless $8\times8$ fermion mass matrix 
$m_{\alpha\beta}\lambda_\alpha \lambda_\beta$. The ${\cal N}=2$ 
$SO(3)$-invariant mass term is a specific choice of $m_{\alpha\beta}$ and 
is given by setting $m_1 =m_2 =m_3 \equiv m$, $m_4=0$ in Eq.\ (\ref{fermion}). 
The branching rules imply that the $\bf 35$ is a counter part of $d=4$, 
fermion masses in $\overline{\bf 10}$ of $SU(4)_R$. 
The ${\cal N}=0$ $SO(3)$-invariant perturbation in four dimensions considered 
in \cite{zamora} consists of a gaugino mass included in the 
$\overline{\bf 10}$ and the $6\times 6$ symmetric traceless scalar mass matrix 
in $\bf 20^\prime$ of $SU(4)_R$. 
The branching rules therefore tell us that the corresponding IIA scalar 
mass term is a $7\times 7$ symmetric traceless mass matrix 
$\mu^2_{ij}\,\chi_i \chi_j$ in $\bf 27$ of $SO(7)_R$. 
The ${\cal N}=0$ $SO(3)$ invariant scalar mass term considered in this paper 
is a specific choice of $\mu^2_{ij}$ and is given by ${\rm Re}\left( 
\mu^2 \phi_i \phi_i \right)$.\\

\noindent
{\bf 3. Dielectric D4-branes wrapping on ${\bf S}^2$}\\

The action for the dielectric D4-branes consists of the Born-Infeld action 
and the Wess-Zumino action and is given by
\begin{eqnarray}
S_{D4} &=& -\mu_4 \int d^5 \xi~ g_s^{-1}Z^{-1/4}\sqrt{ -\det G_\parallel 
\det\left(G_\perp +2\pi\alpha^\prime {\cal F}_2\right) } \nonumber \\ 
&& -\mu_4 \int \left(C_5 +2\pi\alpha^\prime {\cal F}_2 \wedge C_3 \right),
\end{eqnarray}
where $G_\parallel$ stands for the metric along D2-brane world volume 
$\{x^0,x^1,x^2\}$ and $G_\perp$ is a metric on 2-sphere ${\bf S}^2$ in seven 
transverse dimensions. The 2-form field living on the D4-brane is 
$2\pi\alpha^\prime {\cal F}_2 \equiv 2\pi\alpha^\prime F_2 -B_2$.
The D$p$-brane tension is given by $\mu_p =\alpha^{\prime\,(p-1)/2}/
(2\pi\alpha^{\prime})^p$ which reproduces the string tension 
$1/2\pi\alpha^\prime$ when $p=1$. 

Let us introduce the complex coordinates 
$z^i \equiv \frac{1}{\sqrt{2}}(x^{i+2}+ix^{i+6})$ ($i=1,2,3$) 
with $x^6$ as a moduli direction. Suppose that the D2-branes polarize 
into a noncommutative ${\bf S}^2$ under the perturbation, then the ${\bf S}^2$ 
is specified by a single complex coordinate $z$ through $z^i =z e^i$ with 
a real unit vector $e^i$ and its radius is given by $|z|$. 
The metric of the ${\bf S}^2$ couples with the two-form field strength $F_2$ 
which measures the D2-brane charge $n$: 
$$
\int_{{\bf S}^2}F_2 =2\pi n.
$$
The same $F_2$ also arises in the Wess-Zumino action through the term 
${\cal F}_2 \wedge C_3$. If we suppose the large D2-brane charge density 
$n\gg N^{1/2}$, the $F_2$-dependent terms in both Born-Infeld and 
Wess-Zumino action cancel each other and yield a term quartic in $|z|$. 
By taking the Poincar\'{e} dual of the IIA field equation
\begin{equation}
d(*F_4 +B_2 \wedge F_4)=0,\label{sugra1}
\end{equation}
the linear perturbation of $H_3$ and $F_4$ arises in the $C_5$ term of the 
Wess-Zumino action and provides a term cubic in $|z|$. 
Thus, after dividing by the D2-brane world volume $V$, 
we obtain the D2/D4-brane action \cite{bena2}
\begin{equation}
\frac{-S_{D4}}{V}=\frac{2\mu_4}{g_s n\alpha^\prime}
\Biggl[ |z|^4 -2\pi n \alpha^\prime {\rm Im}\left( mzz\bar{z} \right)
+(\pi n \alpha^\prime)^2 m^2 |z|^2 \Biggr],
\label{d2d4}
\end{equation}
which describes the dielectric D4-brane action where $n$ D2-branes polarize 
into the noncommutative ${\bf S}^2$ so that the D4-brane world volume has a 
geometry $R^3 \times\,{\bf S}^2$. Due to the ${\cal N}=2$ supersymmetry, 
the third term in the action was added so as to complete the square and 
to obtain a supersymmetric minimum at $z=i\pi n \alpha^\prime m$. 

Now we move on our main goal of this paper, that is the ${\cal N}=0$ 
$SO(3)$-invariant deformation of 3-dimensional super Yang-Mills, 
by adding $SO(3)$-invariant perturbations which fully breaks ${\cal N}=2$ 
supersymmetry in 3-dimensions. They correspond to the gaugino mass $m_4$ 
in $\bf 35$ and the scalar mass which is given by a traceless $7\times 7$ 
matrix and transforms as $\bf 27$ of $SO(7)_R$ $R$-symmetry. 
In the D4-brane action (\ref{d2d4}), this procedure is achieved by shifting 
\begin{eqnarray*}
{\rm Im}\left( mzz\bar{z} \right) &\longrightarrow& 
{\rm Im}\left( mzz\bar{z}+\frac{m_4}{3}zzz \right), \\
m^2 |z|^2 &\longrightarrow& \left( m^2 +\frac{|m_4|^2}{3} \right)
|z|^2 +{\rm Re}\left( \mu^2 zz \right).
\end{eqnarray*}
For simplicity, we rescale the complex coordinate $z$ and introduce 
the dimensionless parameters $\tilde{b}$ and $c$ such that
\begin{equation}
z=i\pi n \alpha^\prime m\, x e^{i\varphi},\quad
\tilde{b}\equiv -\frac{m_4}{m},\quad c\equiv\frac{\mu^2}{m^2}.
\end{equation}
Then we obtain the D4-brane potential with the $SO(3)$-invariant 
non-supersymmetric perturbation
\begin{equation}
V_{D4}(x,\tilde{b},c) = 2\pi\sqrt{\alpha^{\prime}}
\left(\frac{m^4 n^2}{16\pi}\right)\frac{n}{g_s} x^2 \Biggl[
x^2 -2x {\rm Re}\left(e^{-i\varphi}+e^{3i\varphi}\frac{\tilde{b}}{3}\right)
+1+\frac{|\tilde{b}|^2}{3}-{\rm Re}\left(e^{2i\varphi}c\right)\Biggr],
\end{equation}
which coincides with the D3/D5 potential in \cite{zamora} except for 
the ratio of D-brane tensions $\mu_4 /\mu_5 =2\pi\sqrt{\alpha^{\prime}}$. 
Note that \cite{zamora,polchinski} introduced fermions which transform 
as $\bf 4$ of $SU(4)_R$ $R$-symmetry in $d=4$, ${\cal N}=4$ super Yang-Mills, 
whereas we have used fermions in \cite{bena2} which transform as 
${\bf 3}+{\bf 1}$ of $SU(3)$ $\subset$ $SU(4)_R$.\\

\noindent
{\bf 4. Dielectric NS5-branes wrapping on ${\bf E}^3$}\\

The NS5-brane action with $n$ D2-branes polarized into a 3-ellipsoid 
${\bf E}^3$ has been studied in \cite{bena2} based on the type IIA 
NS5-brane action formulated in \cite{bandos}. 
The action is quite similar to the M5-brane 
action which couples with the self-dual 4-form field strength in M-theory 
\cite{pasti,schwarz}. It consists of the Born-Infeld term, the Wess-Zumino 
term and the mixed term which is necessary to build in the self-dual field 
strength in manifestly 6-dimensional covariant way by invoking a certain 
auxiliary fields. After eliminating the auxiliary field by choosing for 
example $x^2$ as a special direction, the action becomes similar to the 
M5-brane action which shows only 5-dimensional covariance \cite{schwarz} 
and is given by 
\cite{bena2}
\begin{eqnarray}
S_{NS5} &=& S_{BI}+S_{\rm mix}+S_{WZ}, \nonumber\\
S_{BI} &=& -\mu_5 \int d^6 \xi~ g_s^{-2} Z^{-1/2}
\sqrt{ -\det \left( G_{mn}+ig_s Z^{1/4}D_{mn} \right) }, \nonumber\\
S_{\rm mix} &=& -\mu_5 \int d^6 \xi~ \frac{1}{4}
\sqrt{ \frac{-G}{G_{22}} }~ D^{mn}D_{mn2}, \nonumber\\
S_{WZ} &=& -\mu_5 \int \left(B_6 -\frac{1}{2}F_3 \wedge C_3 \right),
\label{d2ns5}
\end{eqnarray} 
where $G$ is a determinant of a 6-dimensional metric $G_{\mu\nu}$, 
$\mu,\nu=0,1,2,3,4,5$ and its 5-dimensional restriction is $G_{mn}$, 
$m,n=0,1,3,4,5$. The 2-form field $D_{mn}$ is given by
$$
D^{mn}=\frac{ \sqrt{G_{22}} }{3!\sqrt{-G}}~\epsilon^{2mnpqr}D_{pqr},
$$
where $D_{pqr}$ is a 5-dimensional component of a 3-form 
$D_3 \equiv F_3 -C_3$. The 6-dimensional self-dual constraints is obtained 
as the equation which determines $D_{mn2}$ in terms of $D_{mn}$ by solving 
the 6-dimensional Euler-Lagrange equations of motion. 

When the 3-ellipsoid is situated in the 3456-plane, the nonzero components 
of $D_3$ are $D_{345}=F_{345}-C_{345}$ and its permutations. 
The 3-form field $F_3$ is determined by the quantization of the D2-brane 
charge along the 3-ellipsoid
$$
\mu_2 \int_{{\bf E}^3} F_3 =2\pi n.
$$
The 3-form potential $C_3$ is given by solving the IIA field equations 
(\ref{sugra1}) and 
\begin{equation}
2d(e^{-2\Phi}*H_3)=F_4 \wedge F_4,\label{sugra2}
\end{equation}
and depends on the fermion mass perturbation (\ref{fermion}) with 
setting $m_1=m_2=m_3\equiv m$ on the D2-branes. The 6-form potential $B_6$ 
in the Wess-Zumino action can be shown to be zero by taking the Poincar\'e 
dual of Eq.\ (\ref{sugra2}). 

In the limit when D2-brane charge is much bigger than NS5-brane charge, 
say $n\gg N^{1/2}$, the action can be expanded with respect to $D_{345}$. 
In this approximation, the Wess-Zumino action is fully given by the 
interaction of the dissolved D2-branes and canceled by those in the 
Born-Infeld and the mixed actions to yield the simplified action \cite{bena2}
\begin{eqnarray}
\frac{-S_{NS5}}{2\pi^2 \mu_5 V} &=&
\frac{3}{16g_s^3 A}\left(3 |z|^4 |w|^2 +|z|^6 \right)
-\frac{1}{4g_s^2}{\rm Re}\left(3m wzz\bar{z}+m_4 zzz\bar{w}\right)
\nonumber \\ &&
+\frac{A}{12g_s}\left(3m^2 |z|^2 +m_4^2 |w|^2 \right), \label{ns5}
\end{eqnarray}
where $A\equiv 4\pi n(\alpha^\prime)^{3/2}$ and $w=x^6$ corresponds to 
the fourth complex coordinate $z^4 =x^6+i x^{10}$ in M-theory. 
The first term of the action is the gravitational energy of the NS5-brane 
and is attractive. The second term is proportional to the linear perturbation 
of $C_{345}$ and is repelling. The balance between the two terms determines 
a finite size 3-ellipsoid. The last term is introduced to complete the square 
in the action in the following sense. 
Since we are interested in the gauge theory on $n$ D2-branes, we give a mass 
$m_4$ only to a gaugino field in the ${\cal N}=1$ gauge multiplet and 
supersymmetry is fully broken. On the other hand, the D2/NS5 bound state in 
the theory is a descendant of an M2/M5 bound state in parent M-theory 
\cite{bena1}. 
The field theory on $n$ M2-branes is a super conformal fixed point of $d=3$, 
${\cal N}=8$ super Yang-Mills and the ${\cal N}=2$ gauge multiplet turns to 
an ${\cal N}=2$ hypermultiplet at the fixed point. 
The $SO(3)$-invariant configuration corresponds to the ${\cal N}=2$ 
supersymmetric configuration in M-theory where three of hypermultiplets have 
the same mass $m$ and the fourth hypermultiplet has a mass $m_4$. 
Although supersymmetry is fully broken by the gaugino mass $m_4$ in IIA 
theory, we can complete the square in the NS5-brane action due to the 
hidden ${\cal N}=2$ supersymmetry of parent M-theory and can find 
supersymmetric minimum at \cite{bena2}
\begin{equation}
z^2 =\frac{2Ag_s}{3}m\sqrt{\frac{m_4}{m}},\quad
x_6^2 =\frac{2Ag_s}{3}m\sqrt{\frac{m}{m_4}}. \label{smini1}
\end{equation}

Let us proceed to the analysis of ${\cal N}=0$ $SO(3)$-invariant deformation, 
which is our main goal in this paper. We introduce the same $SO(3)$-invariant 
scalar mass term, which is in $\bf 27$ of $SO(7)_R$ symmetry, as in the 
D4-brane action. We only have to shift the quadratic term as 
$$
m^2 |z|^2 \longrightarrow m^2 |z|^2 +{\rm Re}\left( \mu^2 zz \right).
$$
In contrast with D2/D4 potential, the mass ratio $m_4 /m$ arises at the 
supersymmetric minimum as the aspect ratio of 3-ellipsoid so that it must be 
always positive. We will therefore use the parameters $b \equiv m_4/m$ 
instead of $\tilde{b}$ itself and the same parameter $c$ as before. 
Rescaling the coordinates such that
\begin{equation}
z^2 =\frac{2Ag_s}{3} m\sqrt{b}\, x^2,\quad
x_6^2 =\frac{2Ag_s}{3} m\frac{1}{\sqrt{b}}\, y^2, \label{rescal}
\end{equation}
the D2/NS5-brane action (\ref{ns5}) becomes
\begin{equation}
U(x,y,b,c)=\frac{A^2 m^3}{18} \sqrt{b} \Biggl[ b x^6 +3 y^2 x^4 
-2(3+b) y x^3 +3(1+c) x^2 +b y^2 \Biggr],
\label{uns5}
\end{equation}
which is a two-dimensional potential depending on two coordinates $x$ and $y$. 
In order to find out the local minima of the potential, we first solve the 
equation $\partial U/\partial y =0$. Then we obtain a trajectory
\begin{equation}
y =\left(\frac{3+b}{3x^4+b}\right)x^3, \label{traj}
\end{equation}
along which the potential remains flat. Substitution of this equation back 
into the potential (\ref{uns5}) gives us one-dimensional potential
\begin{equation}
V_{NS5}(x,b,c)=\frac{A^2 m^3}{18}b^{3/2}
\left[\frac{3x^2}{3x^4+b}\right]
\Biggl[\left(x^4 -1\right)^2+c+\frac{3c}{b}x^4 \Biggr],
\label{vns5}
\end{equation}
of which local minima can be identified with those of the original potential 
(\ref{uns5}). Note that when $c=1$ the potential has a zero at $x=1$ 
corresponding to the supersymmetric minimum of Eq.\ (\ref{smini1}).

\begin{figure}
\noindent
\centerline{\hbox{\psfig{file=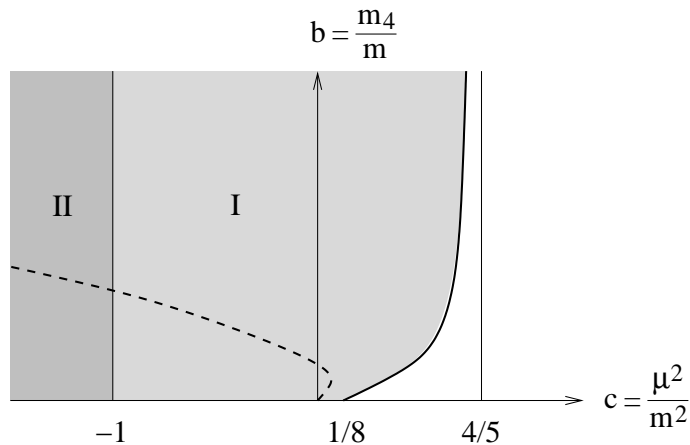,height=7cm}}}\vspace{-4mm}
\caption{\sl The phase diagram of D2/NS5-branes. 
The thick line is the critical line $c=h(b)$ for a D2/NS5-minimum without 
D4-brane charge. The allowed region in the left hand side of the critical 
line is separated into phase I and phase II. The local maximum disappears 
in phase II. The dashed line is the critical line for a D2/NS5-branes in 
the presence of D4-flux which will be discussed in Section 5.}\label{ns5ph} 
\end{figure}

Let us find out the regions in $(b,c)$-plane where we have a finite size 
3-ellipsoid. Differentiating the potential (\ref{vns5}) with respect to $x$, 
one finds that the equation which determines local minima of the potential 
is given by the cubic equation
\begin{equation}
f(X)\equiv X^3 + c_1 X^2 +c_2 X +c_3 =0, \label{cubic}
\end{equation}
where $X\equiv x^4$ and $c_1$, $c_2$, and $c_3$ are determined as
$$
c_1=\frac{5b^2-6b+9c}{9b},\quad c_2=-\frac{2b-2c+1}{3},
\quad c_3=\frac{b\left(1+c\right)}{9}.
$$
The extrema of the cubic function $f(X)$ are located at
\begin{equation}
X_\pm =\frac{1}{3}\left(-c_1 \pm \sqrt{c_1^2 -3c_2}\right). \label{extrema}
\end{equation}
Since $c_1^2 -3c_2$ takes positive values, we always have two extrema. 
Furthermore, we can easily see that $X_-$ is always negative in the whole of 
$(b,c)$-plane. The allowed region of $(b,c)$-plane, where the cubic equation 
(\ref{cubic}) has at least one solution, is therefore given by the inequality
\begin{equation}
f(X_+) \le 0 \longleftrightarrow 
\left( 2c_1^3 -9c_1 c_2 +27 c_3 \right)^2 \le 
4\left( c_1^2 -3c_2 \right)^3, \label{ineq}
\end{equation}
which can be solved with respect to $c$ such that
\begin{equation}
c\le h(b)\equiv\frac{1}{216}\left[ 9+150b+97b^2 +H(b)^{1/3}
+\frac{(3+b)^2 \left(9+6b-2591b^2\right)}{H(b)^{1/3}} \right], \label{smallh}
\end{equation}
where the function $H(b)$ is given by
$$
H(b) \equiv (3+b)^2 \left[81+108b+58374b^2+38892b^3-833327b^4
+144b\left(9+6b+325b^2\right)^{3/2}\right].
$$

The corresponding phase diagram is shown in Fig.\ \ref{ns5ph}. 
Since the parameter $b$ turns to the aspect ratio of 3-ellipsoid 
in the supersymmetric limit $c \to 0$ and therefore should be positive, 
the negative half of $b$ axis is forbidden. 
The critical line $c=h(b)$ approaches to $4/5$ when $b \to \infty$ which is 
in perfect agreement with the upper-bound of $c$ for the D3/NS5 bound state 
in \cite{zamora}. The critical line intersects the $c$ axis at $c=1/8$ which 
is also the same as in the D3/NS5 bound state. The cubic function in 
Eq.\ (\ref{cubic}) has two extrema corresponding to a local maximum and 
a local minimum of the potential (\ref{vns5}) when $c_3>0$ ($c>-1$), 
whereas it has only one minimum when $c_3 <0$ ($c<-1$). 
Hence the allowed region in $(b,c)$-plane was separated into phase I 
with $c>-1$ and phase II with $c<-1$. Again, the critical line $c=-1$ 
coincides with the upper-bound of the region where two D3/NS5 minima coexist 
in \cite{zamora}. 

\begin{figure}
\centerline{\hbox{\psfig{file=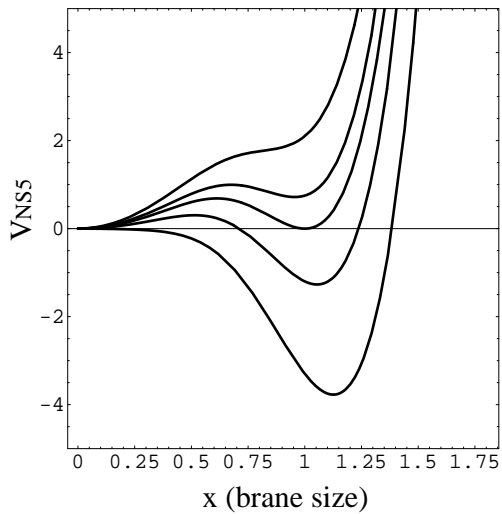,width=7cm}}}
\caption{\sl The D2/NS5 potential (\ref{vns5}) when $b=2$ with varying $c$. 
The horizontal axis $x$ denotes the rescaled S$^2$ radius (\ref{rescal}) 
of a 3-ellipsoidal shell for a D2/NS5-brane. Each line corresponds to 
$c=0.6$, $0.25$, $0$, $-0.4$, and $-1.1$ in order from above.}
\label{poten}
\end{figure}

Now let us look at some aspects of the potential (\ref{vns5}) with 
varying $c$ for a fixed value of $b$. We set $b=2$ for simplicity so that 
the critical value of $c$ is given by $c^* =h(b=2)\approx 0.556281$. 
In Fig.\ \ref{poten}, five aspects of the potential are depicted. 
Each line corresponds to $c=0.6$, $0.25$, $0$, $-0.4$, and $-1.1$ in order 
from above. When $c>c^*$ the potential has no local minima except for the 
origin and therefore any finite size ellipsoid does not exist (the first line 
with $c=0.6>c^*$ in Fig.\ \ref{poten}). We find a local minimum 
other than the origin when $c=0.25$ though the potential has positive energy 
at the point. In the supersymmetric limit $c=0$, we find a local minimum at 
$x=1$ and the potential energy becomes zero as required by supersymmetry. 
When the parameter $c$ becomes negative (the fourth line with $c=-0.4$ 
in Fig.\ \ref{poten}), the potential energy at the finite 
size local minimum turns to negative so that we can identify the minimum 
point as a stable finite size solution. Finally, if the parameter $c$ becomes 
smaller than $-1$ and enters into phase II (the fifth line with 
$c=-1.1$ in Fig.\ \ref{poten}), the local maximum point 
disappears. The trivial solution at the origin becomes unstable and the 
vacuum necessarily goes to formation of the finite size 3-ellipsoid.\\

\noindent
{\bf 5. Dielectric NS5-branes wrapping on ${\bf E}^3$ with D4-brane charge}\\

The general action for NS5-branes in the type IIA theory was found recently 
in \cite{bandos}. The Wess-Zumino term of the action contains, 
other than the NS-NS six-form potential $B_6$, the coupling between the bulk 
R-R five-form potential $C_5$ and the one-form field strength $\cal F$ living 
on the NS5-brane. Since a nonzero $\cal F$ corresponds to a nonzero D4-brane 
charge, we have to take it into account when we analyze D2/D4/NS5 bound states 
and its action is obtained by shifting $B_6 \longrightarrow B_6 +C_5 \wedge 
{\cal F}$ in the Wess-Zumino term of the D2/NS5-brane action (\ref{d2ns5}) 
to yield
\begin{equation}
\delta S_{WZ}=-\mu_5 \int C_5 \wedge {\cal F}.\label{d4wz}
\end{equation}

Let us derive the D2/D4/NS5-brane action explicitly from the D2/NS5-brane 
action \cite{bena3}. First, the nonzero D4-brane charge may possibly cause 
a rotation of the 3-ellipsoid with an aspect ratio $\alpha$ in 3-7, 4-8, 5-9 
planes at an angle $\gamma$. This rotation is achieved by setting 
$z=re^{i\gamma}$, and $w=\alpha r$ in the D2/NS5-brane action (\ref{ns5}) 
to obtain
\begin{eqnarray}
\frac{-S_{NS5}}{2\pi^2 \mu_5 V} &=& \frac{3r^6}{16g_s^3 A}
\left(3\alpha^2 +1\right)
-\frac{\alpha r^4}{4g_s^2}\left(3m\cos\gamma +m_4 \cos 3\gamma\right)
\nonumber \\ &&
+\frac{Ar^2}{12g_s}\left(3m^2 +\alpha^2 m_4^2 +3\mu^2 \cos 2\gamma\right),
\end{eqnarray}
where the $SO(3)$-invariant scalar mass $\mu^2$ is accompanied by 
a factor of $\cos 2\gamma$ induced by the rotation. Then, we evaluate the 
D4-brane charge contribution (\ref{d4wz}) on the 3-ellipsoid and minimize it 
to obtain \cite{bena3}
$$
\frac{-\delta S_{WZ}}{2\pi^2 \mu_5 V}=
-\frac{A r^2}{4g_s}\left(m\sin\gamma+\frac{m_4}{3}\sin3\gamma\right)^2.
$$
Finally, the generalized NS5-brane action is given by
$$
S_{GNS5}\equiv S_{NS5}+\delta S_{WZ},
$$
which still has a supersymmetric minimum of (\ref{smini1}) at $\gamma=0$ 
when $c=0$. Again, rescaling the coordinates such that
\begin{equation}
r^2 =\frac{2Ag_s}{3}m \sqrt{b}\,x^2,\quad
\alpha r^2 =\frac{2Ag_s}{3}m \frac{1}{\sqrt{b}}\,y^2, \label{rescal2}
\end{equation}
the action $S_{GNS5}$ turns into the D2/D4/NS5-brane potential
\begin{eqnarray}
U(x,y,\gamma,b,c) &=& \frac{A^2 m^3}{18} \sqrt{b} \Biggl[ b x^6 +3 y^2 x^4 
-2(3\cos\gamma +b\cos 3\gamma) y x^3 \nonumber \\ &&
+3\biggl[ 1+c \cos 2\gamma
-\biggl(\sin\gamma +\frac{b}{3}\sin 3\gamma\biggr)^2 \biggr] x^2 
+b y^2 \Biggr]. \label{ud4ns5}
\end{eqnarray}
which reproduces the D2/NS5-brane potential (\ref{uns5}) when $\gamma=0$ as 
expected.

We proceed to the analysis of local minima of the potential in 
three-dimensional coordinate space $(x,y,\gamma)$. The trajectory along which 
the $y$-derivative of the potential is always zero is given by
\begin{equation}
y=\left(\frac{3\cos\gamma+b\cos 3\gamma}{3x^4+b}\right)x^3,
\end{equation}
which reproduces Eq.\ (\ref{traj}) when we turn off the D4-brane charge 
($\gamma=0$). Substitution of this equation back into the potential 
(\ref{ud4ns5}) provides two-dimensional potential
\begin{eqnarray}
V_{GNS5}(x,u,b,c) &=& \frac{A^2 m^3}{18}b^{3/2}
\left[\frac{3x^2}{3x^4+b}\right]
\Biggl[\left(x^4 -u\right)^2 +cu +\frac{3cu}{b}x^4 \nonumber \\ &&
+\frac{(2u+1)(u-1)}{18}\left[(2u+1)b^2+6b-9\right] \Biggr], \label{vd4ns5}
\end{eqnarray}
where $u\equiv \cos 2\gamma$ was introduced as a new coordinate. 
We notice that the potential (\ref{vd4ns5}) coincides with the D2/NS5-brane 
potential (\ref{vns5}) when we turn off the D4-brane charge. 
In contrast with the D2/NS5 case, we have to solve both $x$- and 
$u$-flatness conditions in order to determine the allowed region for the 
stable D2/D4/NS5 minima in the $(b,c)$-plane. 

\begin{figure}
\centerline{\hbox{\psfig{file=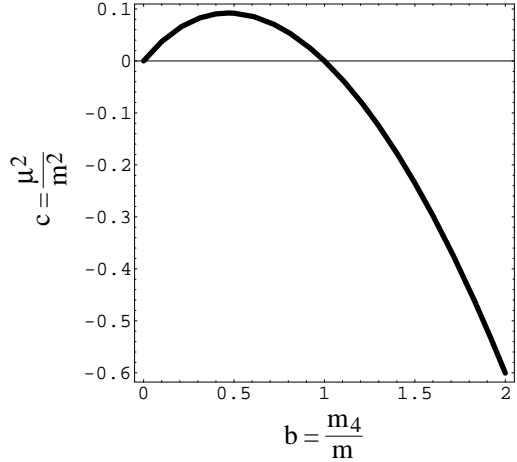,width=7cm}}}
\caption{\sl The critical line (a dashed line in Fig.\ \ref{ns5ph}) 
for a stable D2/NS5 minimum $(u=1)$ in the presence of D4-flux. 
The maximum point is at $(b,c)\approx(0.468,0.0922)$ which 
is below the critical line (a thick line in Fig.\ \ref{ns5ph}) 
for a D2/NS5 minimum without D4-brane charge. 
The line intersects the $b$-axis at $(b,c)=(0,0)$ and $(1,0)$, and goes 
down to $c\approx -0.601$ at $b=2$.}\label{bc01} 
\end{figure}

The $x$-flatness condition is given by the same cubic equation (\ref{cubic}) 
as before except that its coefficients are modified to be
\begin{eqnarray*}
c_1 &=& \frac{5b^2 -6bu+9cu}{9b},\\
c_2 &=& -\frac{1}{54}\left[ (2u+1)^2 (u-1) b^2 +6(2u^2+5u-1)b
+9(1+u)-36cu \right],\\
c_3 &=& \frac{b}{162} \Biggl[ (2u+1)(u-1)\left[ (2u+1)b^2+6b \right]
+9(1+u+2cu)\Biggr].
\end{eqnarray*}
We again see that $c_1^2 -3c_2$ takes positive values and $X_- <0< X_+$ 
so that the condition which restricts the allowed 
region for finite size minima is given by the same inequality (\ref{ineq}) 
as before. The inequality is solved with respect to $cu$ to yield $cu \leq 
h(b,u)$ which coincides with the inequality (\ref{smallh}) when D4-brane 
charge is turned off at $u=1$ ($\gamma=0$). 
However, the function $h(b,u)$ is no longer 
bounded from above for generic values of $u\neq 1$. This simply means that 
the $x$-flatness is not enough to determine the local minima of the potential 
once we turn on the D4-brane charge. The correct procedure is first we solve 
the cubic equation (\ref{cubic}) and find its largest solution $X_{\rm max}$ 
corresponding to the potential minimum along $x$-axis, then substitute it 
into the $u$-flatness condition $(\partial/\partial u)V_{GNS5}=0$ to yield
\begin{equation}
X_{\rm max}=-\frac{b\Bigl[b^2 (1-4u^2)+2b(1-4u)-3b(1+2c)\Bigr]}{6(2b-3c)},
\label{d2d4ns5}
\end{equation}
which determines the critical surface $g(u,b,c)=0$ giving a foliation 
of trajectories of a stable D2/D4/NS5 minimum in $(u,c)$-plane along $b$.

\begin{figure}
\centerline{\hbox{\psfig{file=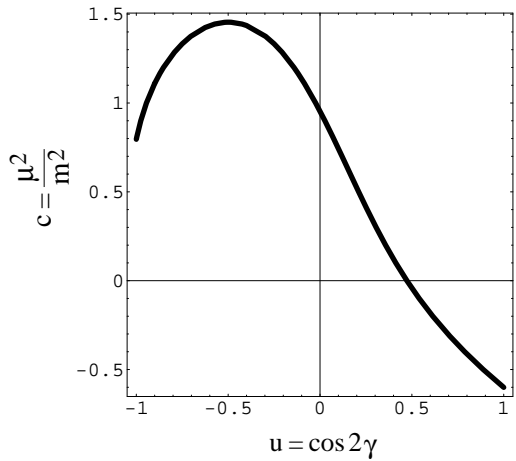,width=7cm}}}
\caption{\sl The trajectory of a D2/D4/NS5 minimum when $b=2$. 
The D2/NS5 minimum continues to exist even under non-zero D4-flux 
when $c<c_- \approx-0.601$. At $c=c_-$, the trajectory starts to move along 
$u$-direction to yield a D2/D4/NS5 minimum. Finally, the minimum disappears 
at $(u,c)\approx(-0.50,1.455)$.}\label{uc02} 
\end{figure}

\begin{figure}
\begin{center}
\begin{minipage}[t]{6.8cm}
\centerline{\hbox{\psfig{file=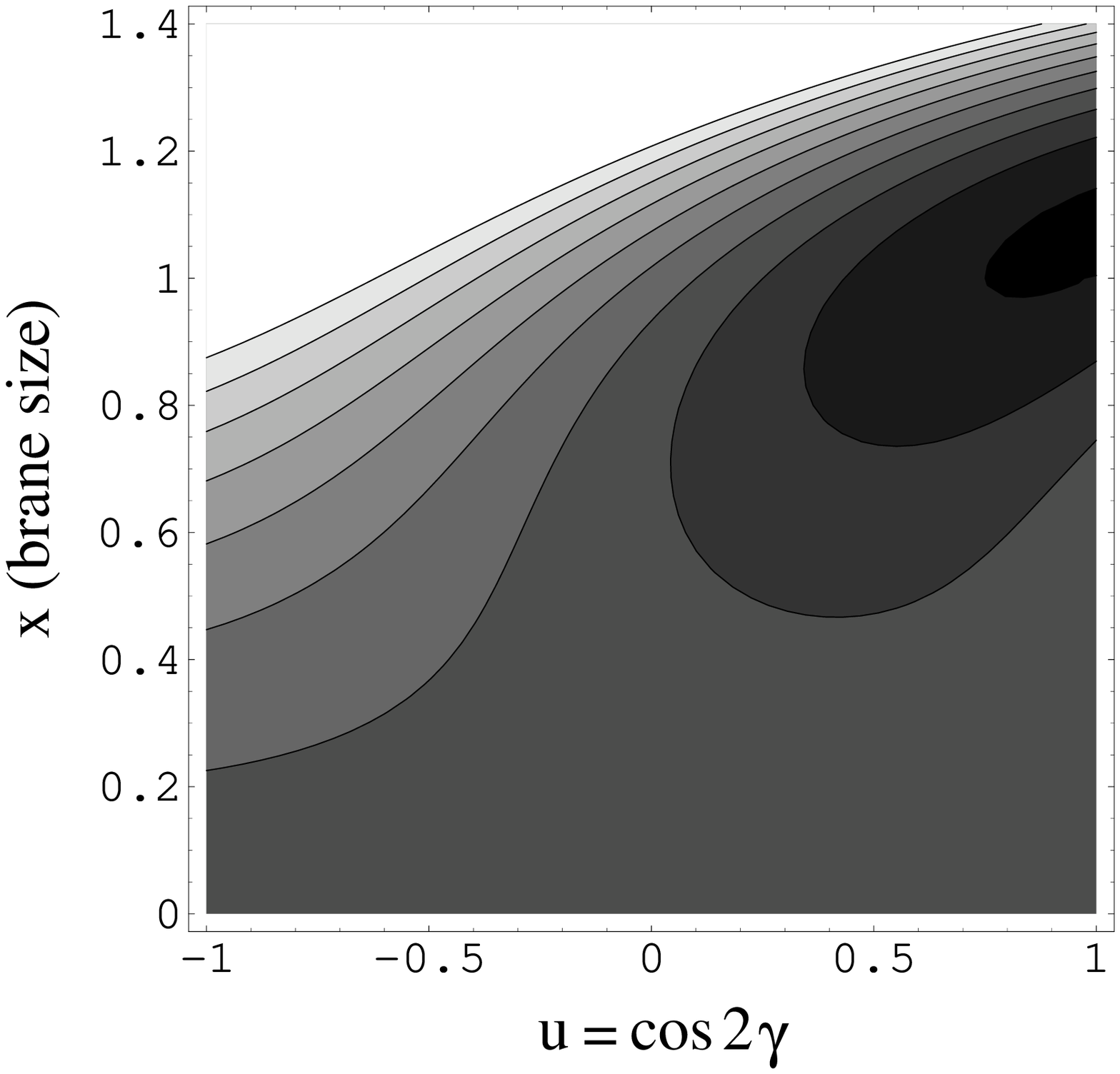,width=6.8cm}}}
\caption{\sl The D2/D4/NS5 minimum when $(b,c)=(2,-0.601)$. It ceases 
to be a D2/NS5 minimum and starts to move along $u$-direction.}\label{cont02}
\end{minipage}\hspace{1cm}
\begin{minipage}[t]{6.8cm}
\centerline{\hbox{\psfig{file=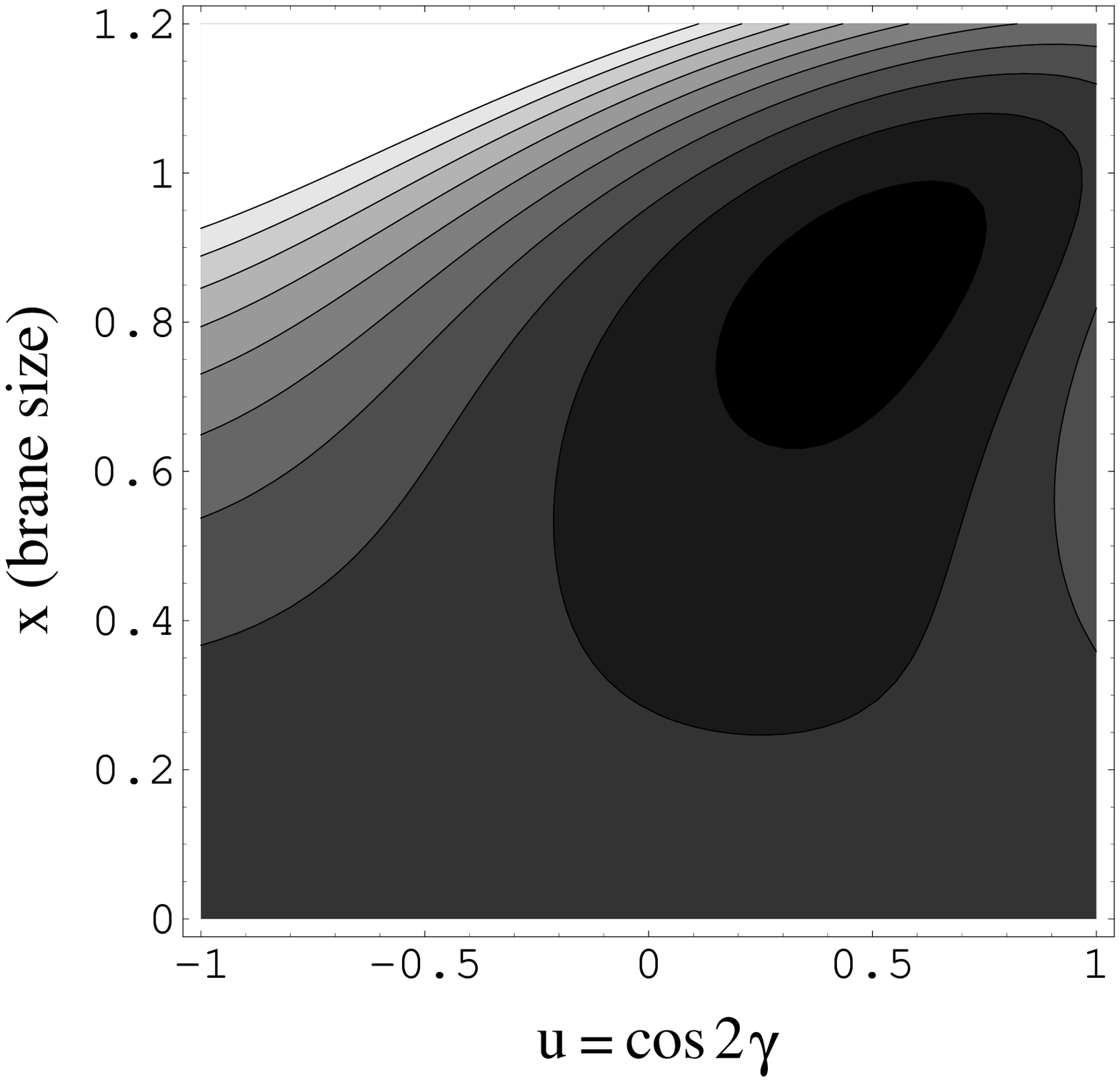,width=6.8cm}}}
\caption{\sl The D2/D4/NS5 minimum when $(b,c)=(2,0)$. A stable minimum 
appears away from $u=1$ and becomes a D2/D4/NS5-minimum.}
\label{cont03}
\end{minipage}
\end{center}
\end{figure}

\begin{figure}
\begin{center}
\begin{minipage}[t]{6.8cm}
\centerline{\hbox{\psfig{file=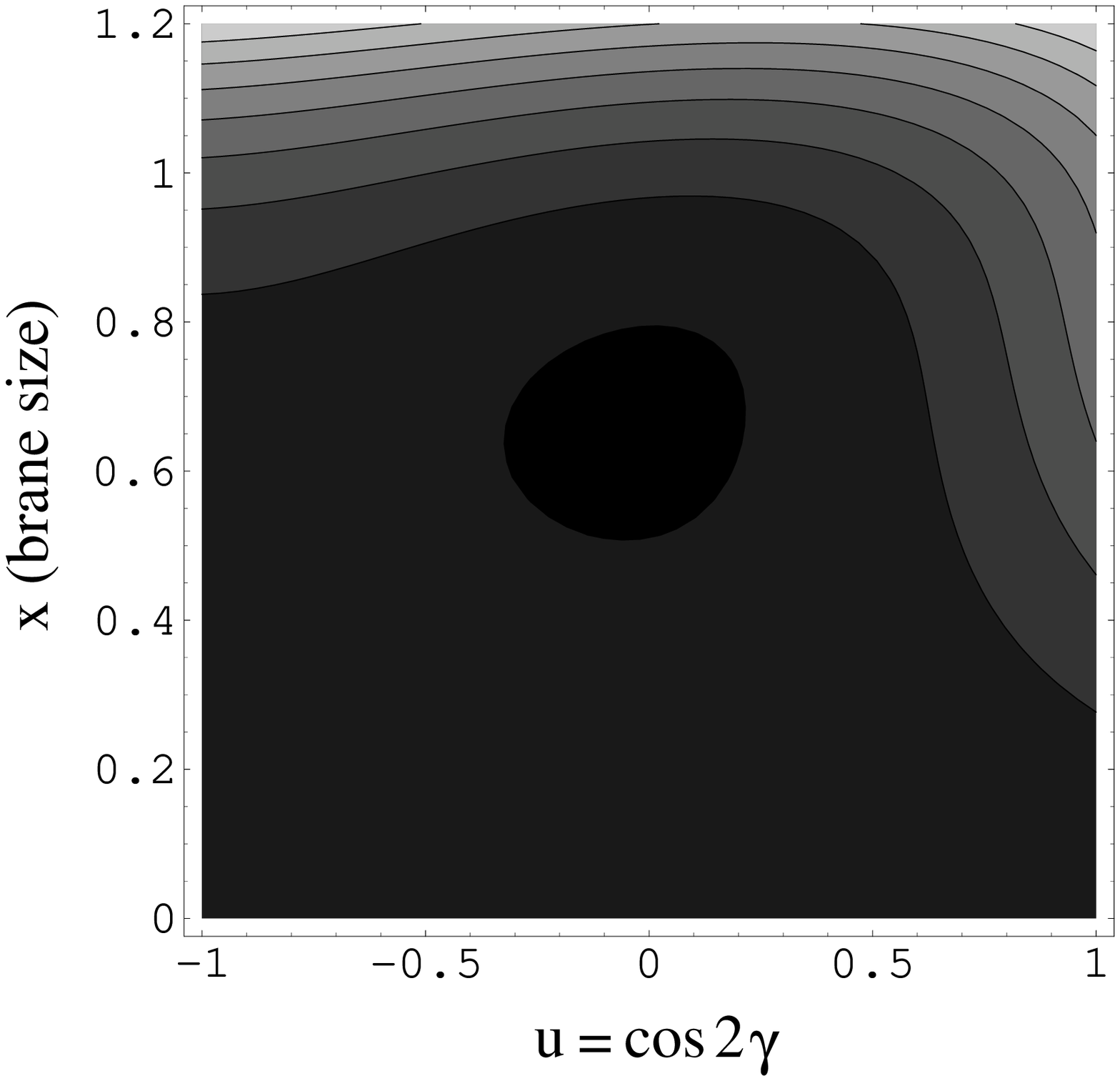,width=6.8cm}}}
\caption{\sl The D2/D4/NS5 minimum when $(b,c)=(2,1)$. 
The minimum approaches the ending point $(u,x)\approx(-0.50,0.71)$ 
which is evident in Fig.\ \ref{cont06}.}\label{cont05}
\end{minipage}\hspace{1cm}
\begin{minipage}[t]{6.8cm}
\centerline{\hbox{\psfig{file=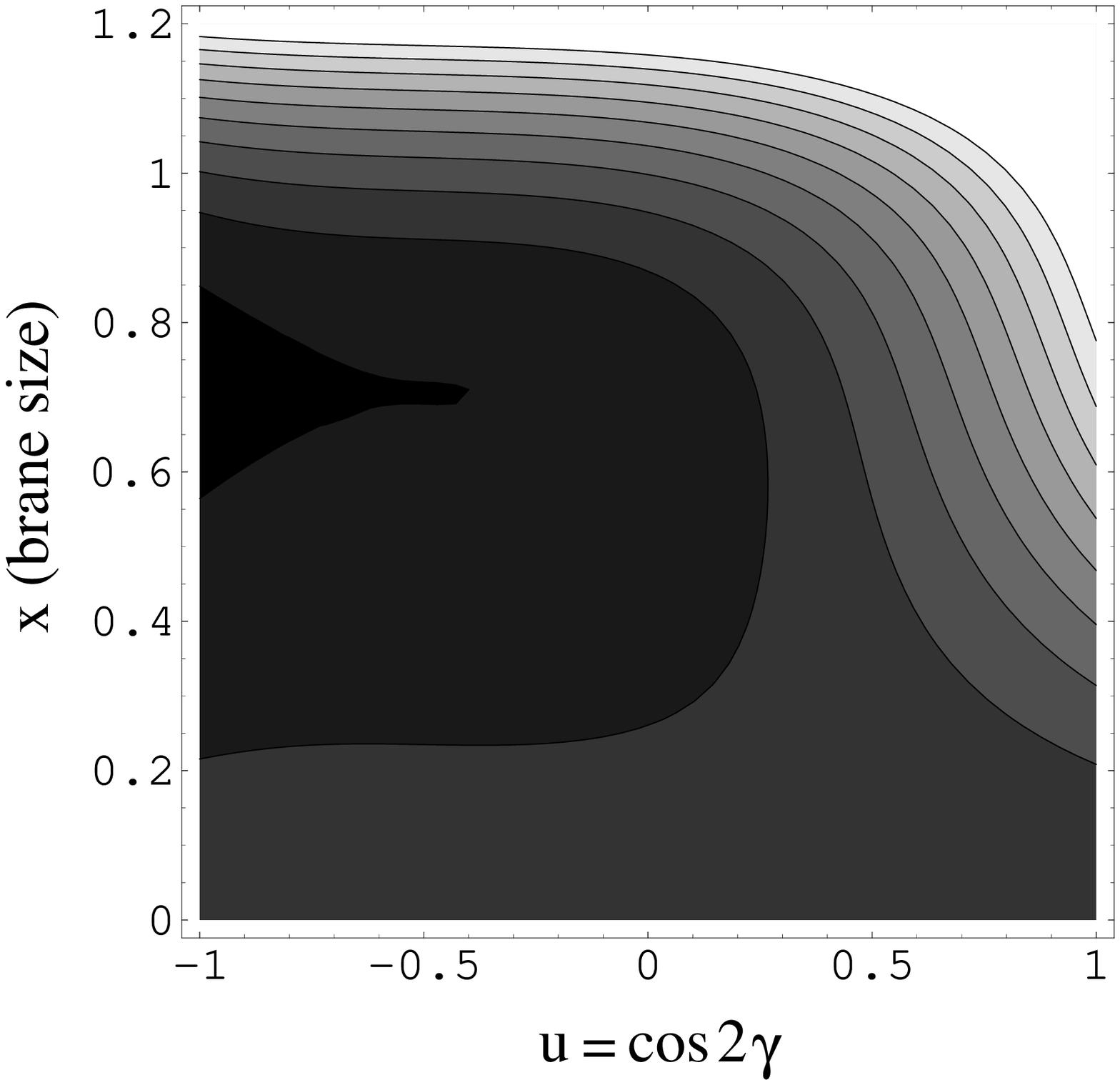,width=6.8cm}}}
\caption{\sl The D2/D4/NS5 minimum when $(b,c)=(2,1.455)$. 
The minimum merges into the $u=-1$ edge and disappears.}\label{cont06}
\end{minipage}
\end{center}
\end{figure}

Let us analyze the D2/D4/NS5 constraint (\ref{d2d4ns5}) precisely. 
The cross section of the surface at $u=1$, say $g(1,b,c)=0$, provides a new 
critical line for a stable D2/NS5 minimum in the presence of D4-flux. 
The new critical line denoted as a dashed line is located in the left hand 
side of the critical line denoted as a thick line for a D2/NS5-minimum 
without D4-brane charge as shown in Fig.\ \ref{ns5ph}. 
The numerical plot of that line is again shown in Fig. \ref{bc01}. 
The maximum point is at $(b,c)\approx(0.468,0.0922)$ which is below 
the critical line for a D2/NS5 minimum without D4-brane charge. 
The line intersects the $b$-axis at $(b,c)=(0,0)$ and $(1,0)$, and goes down 
to $c=c_- \approx -0.601$ at $b=2$. 
To look at the upper and lower bounds of $c$ 
for a specific value of $b$, let us choose for example $b=2$ as before. 
The constraint is now $g(u,2,c)=0$ and provides the trajectory of a D2/D4/NS5 
minimum in $(u,c)$-plane as shown in Fig.\ \ref{uc02}. As we go up along the 
D2/NS5 line $u=1$ from infinitely below, the D2/NS5 minimum continues to stay 
on $u=1$ even under non-zero D4-flux. When $c$ reaches the lower 
critical value of $c_- \approx -0.601$, the trajectory starts to move along 
$u$-direction so that we necessarily have a D2/D4/NS5 minimum. 
Finally, at the maximum point $(u,c)\approx (-0.50,1.455)$, the minimum 
disappears to yield the upper critical value of $c_+ \approx 1.455$. 
We can also demonstrate the behavior of D2/D4/NS5 minimum by using the 
contour plots of the potential (\ref{vd4ns5}) with $b=2$. 
In Fig.\ \ref{cont02}, we can see that the stable D2/NS5 ground state with the 
minimum radius $x\approx1.077$ ceases to stay on $u=1$ and begins to shift 
in the D4-brane charge direction. Then, the D2/D4/NS5 minimum continues moving 
to the ending point at $(u,x)\approx(-0.50,0.71)$ (see Fig. \ref{cont03}, 
and Fig.\ \ref{cont05}) and finally merges into the $u=-1$ edge and 
disappears as shown in Fig. \ref{cont06}.

Though the above analysis is just for $b=2$, it seems to reflect the generic 
feature of D2/D4/NS5 minimum. As we increase the value of $b$, the ending 
point of D2/D4/NS5 minimum approaches to $u=0$ and the upper- and lower-bounds 
in $c$ blow up as $b$ goes to infinity.\\

\noindent
{\bf 6. Conclusions}\\

We have studied the type IIA dielectric D4- and NS5-branes in the bulk 
corresponding to the $SO(3)$-invariant ${\cal N}=2,0$ deformations of 
three-dimensional ${\cal N}=8$ super Yang-Mills. Specifically, we added 
the $SO(3)$-invariant fermion masses in $\bf 35$ of $SO(7)_R$ in ${\cal N}=2$ 
case, and the gaugino mass in $\bf 35$ as well as the scalar masses in 
$\bf 27$ of $SO(7)_R$ in the ${\cal N}=0$ case. We find that the D2/D4 bound 
states show exactly same phase structure as in the D3/D5 bound states in 
type IIB theory as expected from $T$-duality. Moreover, the D2/NS5 bound 
states show the same critical values $c=4/5$, $1/8$ and $-1$ as those for 
D3/NS5 bound states although the critical lines are deformed. This seems 
to be consistent with $T$-duality between IIA and IIB confining vacua. 
We have also examined the D2/D4/NS5 bound states and find the phase for a 
stable D2/NS5 minimum in the presence of D4-flux. The corresponding 
critical line is obtained as a $u=1$ cross section of the critical surface 
$g(u,b,c)=0$. If we fix the parameter $b=2$ as a specific value, the surface 
provides the trajectory of a D2/D4/NS5 minimum in $(u,c)$-plane. 
The minimum continues staying on the D2/NS5 axis $u=1$ until the parameter 
$c$ reaches to the critical value given by $g(1,b,c_- (b))=0$, then it starts 
to shift along $u$-direction to become a D2/D4/NS5 minimum. 
Finally, the minimum disappears at the maximum point of the trajectory when 
$c$ gets to the upper critical value determined by 
$g(u_{\rm min},b,c_+ (b))=0$ and 
$(\partial g/\partial u)(u_{\rm min},b,c_+ (b))=0$. 
This suggests that in gauge theory side oblique confining vacua may exist in 
some particular regions of the scalar mass ${\rm Re}\left(\mu^2 \phi_i \phi_i 
\right)$ bounded from above and below for a given gaugino mass $m_4$. 
There exist many supersymmetric or non-supersymmetric vacua preserving a
particular symmetry in the four-dimensional gauged
supergravity \cite{ahnetal}. It would be interesting to study
corresponding dual gauge theory side by looking at the perturbations in
the supergravity side.\\


C.A. was supported by Korea Research Foundation Grant (KRF-2000-003-D00056). 
T.I. was supported by the grant of Post-Doc.~Program, Kyungpook 
National University (2000).


\newpage

\begin{thebibliography}{99}

\bibitem{agmoo}
O. Aharony, S.S. Gubser, J. Maldacena, H. Ooguri, and Y. Oz, 
Phys. Rep. {\bf 323} (2000) 183 ({\tt hep-th/9905111}).

\bibitem{polchinski}
J. Polchinski and M. Strassler, {\tt hep-th/0003136}.

\bibitem{myers}
R.C. Myers, J. High Energy Phys. {\bf 9912} (1999) 022 ({\tt hep-th/9910053}).

\bibitem{bena1}
I. Bena, Phys. Rev. D {\bf 62} (2000) 126006 ({\tt hep-th/0004142}).

\bibitem{bena2}
I. Bena and A. Nudelman, Phys. Rev. D {\bf 62} (2000) 086008 
({\tt hep-th/0005163}).

\bibitem{bena3}
I. Bena and A. Nudelman, Phys. Rev. D {\bf 62} (2000) 126007 
({\tt hep-th/0006102}).

\bibitem{zamora}
F. Zamora, J. High Energy Phys. {\bf 0012} (2000) 021 ({\tt hep-th/0007082}).

\bibitem{horowitz}
G. Horowitz and A. Strominger, Nucl. Phys. {\bf B360} (1991) 197.

\bibitem{itzhaki}
N. Itzhaki, J. Maldacena, J. Sonnenschein, and S. Yankielowicz, 
Phys. Rev. D {\bf 58} (1998) 046004 ({\tt hep-th/9802042}).

\bibitem{seiberg}
N. Seiberg, Nucl. Phys. Proc. Suppl. {\bf 67} (1998) 158 
({\tt hep-th/9705117}). 

\bibitem{vafa}
C. Vafa and E. Witten, Nucl. Phys. {\bf B431},(1994) 3 
({\tt hep-th/9408074}).

\bibitem{donagi}
R. Donagi and E. Witten, Nucl. Phys. {\bf B460} (1996) 299 
({\tt hep-th/9510101}).

\bibitem{bandos}
I. Bandos, A. Nurmagambetov, and D. Sorokin, Nucl. Phys. {\bf B586} 
(2000) 315 ({\tt hep-th/0003169}).

\bibitem{pasti}
P. Pasti, D. Sorokin, and M. Tonin, Phys. Lett. {\bf B398} (1997) 41 
({\tt hep-th/9701037}).

\bibitem{schwarz}
M. Perry and J. H. Schwarz, Nucl. Phys. {\bf B489} (1997) 47 
({\tt hep-th/9611065});\\
J. H. Schwarz, Phys. Lett. {\bf B395} (1997) 191 ({\tt hep-th/9701008}).

\bibitem{ahnetal}
C. Ahn and S.-J. Rey, Nucl. Phys. {\bf B572} (2000) 188 
({\tt hep-th/9911199});\\
C. Ahn and J. Paeng, Nucl. Phys. {\bf B595} (2001) 119 
({\tt hep-th/0008065});\\
C. Ahn and K. Woo, Nucl. Phys. {\bf B599} (2001) 83 
({\tt hep-th/0011121}).

\end{thebibliography}
\end{document}